\documentclass{aa}
\usepackage[colorlinks=true,linkcolor=blue,citecolor    = blue]{hyperref}
\usepackage{graphicx}
\usepackage{txfonts}
\usepackage{float}
\usepackage{rotating,tabularx}
\usepackage{multirow}
\usepackage{afterpage}
\usepackage{bigstrut}
\usepackage[flushleft]{threeparttable }
\usepackage{amsmath}
\usepackage{enumitem}
\usepackage{amsfonts,color}
\usepackage{amssymb,float}
\usepackage[utf8]{inputenc}
\usepackage{color}
\usepackage{etoolbox}
\usepackage{appendix}
\usepackage{url}
\usepackage{dirtytalk}
\usepackage{float}
\usepackage{fontawesome}

\usepackage{booktabs}
\usepackage[graphicx]{realboxes}
\usepackage{adjustbox}
\usepackage{multirow}
\usepackage{rotating,tabularx}
\usepackage{afterpage}
\usepackage{tablefootnote}

\usepackage{tikz}

\usepackage{bigstrut}

\definecolor{lime}{HTML}{A6CE39}
\DeclareRobustCommand{\orcidicon}{%
    \begin{tikzpicture}
    \draw[lime, fill=lime] (0,0) 
    circle [radius=0.16] 
    node[white] {{\fontfamily{qag}\selectfont \tiny ID}};
    \draw[white, fill=white] (-0.0625,0.095) 
    circle [radius=0.007];
    \end{tikzpicture}
    \hspace{-2mm}
}

\foreach \x in {A, ..., Z}{%
    \expandafter\xdef\csname orcid\x\endcsname{\noexpand\href{https://orcid.org/\csname orcidauthor\x\endcsname}{\noexpand\orcidicon}}
}
\newcommand{\orcid}[1]{\href{https://orcid.org/#1}{\textcolor[HTML]{A6CE39}{\orcidicon}}}
\usepackage{graphicx}

\usepackage{txfonts}

\newcommand{\gaia}{\textit{Gaia}}
\newcommand{\hst}{HST}
\defcitealias{2024arXiv240208843C}{Paper I}
\defcitealias{2021MNRAS.505.5978V}{VB21}
\defcitealias{harris2010new}{H10}
\defcitealias{2014ApJ...787L..26F}{F14}

\begin{document}

\title{Relation between
the geometric shape and \\ rotation of Galactic globular clusters\thanks{Tables \ref{tab:ellipticities} and \ref{tab:position_angles} are available in electronic form at the CDS via anonymous ftp to cdsarc.cds.unistra.fr (130.79.128.5)
or via https://cdsarc.cds.unistra.fr/cgi-bin/qcat?J/A+A/}}

\titlerunning{Geometric shape of Galactic globular clusters}

   \author{Mauricio Cruz Reyes \inst{1}\orcid{0000-0003-2443-173X}
          \and Richard I. Anderson\inst{1}\orcid{0000-0001-8089-4419}}

   \institute{Institute of Physics, \'Ecole Polytechnique F\'ed\'erale de Lausanne (EPFL), Observatoire de Sauverny, Chemin Pegasi 51b, 1290 Versoix, Switzerland  \\
    \email{mauricio.cruzreyes@epfl.ch, richard.anderson@epfl.ch}   }

   \date{Received \today}

  \abstract
  {We homogeneously measured the elliptical shapes of 163 globular clusters (GCs) using the on-sky distribution of their cluster members and the third data release of the ESA mission \gaia\ (DR3).  The astrometry enables the differentiation of stars within clusters from those in the field.  This feature is particularly valuable for clusters located in densely populated areas of the sky, where conventional methods for measuring the geometry of the GCs are not applicable.  The median axial ratio of our full sample is $\langle b/a \rangle =  0.935^{+0.033}_{-0.090}$ and $0.986^{+0.009}_{-0.004}$ for the subset of 11 GCs previously studied based on \textit{Hubble} Space Telescope imaging.  We investigated whether the minor axis of the ellipses can be interpreted as a pseudo-rotation axis by comparing it to measurements of cluster rotation.   Using the radial velocities from \gaia, we detected rotation for three clusters, NGC~5139, NGC~104, and NGC~6341, and observed an alignment between the pseudo-rotation axis and the 2D projection of the real rotation axis. To expand the set of clusters for which rotation has been detected, we analyzed multiple literature references. Depending on the reference used for comparison, we observed an alignment in between 76\% to 100\% of the clusters. The lack of an alignment observed in some clusters may be linked to different scales analyzed in various studies. Several studies have demonstrated that the orientation of rotation varies with the distance from the center.  We estimate that the next \gaia\ release will increase the number of stars with radial velocities in GCs from  $\sim 10,000$ in \gaia\ DR3 to $\sim 55,000$ in \gaia\ DR4. This will enable the measurement of rotation and ellipticities at identical angular scales for additional clusters, which will help us to clarify whether the previously mentioned alignment occurs in all clusters.}


   \maketitle

\section{Introduction}
Globular clusters (GCs) are often thought of as spherical distributions of stars by nonspecialists. Nonetheless, the first reports of nonsphericity date back to more than a century ago, when \citet{1917PNAS....3...96P} examined the distribution of stars in five GCs that were observed using photographic plates. The authors concluded that ellipsoids provided a more suitable approximation of their real geometry. This insight gained further support from \citet[henceforth: H10]{harris2010new}, who compiled parameters for 157 GCs and showed that deviations from sphericity are rather common: 94 of the 157 GCs exhibited them. However, the quality of these measurements was somewhat limited by the spatial resolution of the observations, which primarily came from the catalog of \citet{1987ApJ...317..246W}, which used the Palomar \citep{1991PASP..103..661R} and the Southern Sky survey \citep{1974A&AS...18..463H} photographic plates.   Nevertheless, \citet[henceforth: C10]{2010ApJ...721.1790C} provided further evidence of nonsphericity by analyzing the spatial distribution of sources with Two Micron All-Sky Survey (2MASS) photometry \citep{2006AJ....131.1163S}, revealing similar deviations in 116 clusters, although this time with CCD measurements. Using the astrometry from the Early Data Release 3 (EDR3) of \gaia, \citet{2021MNRAS.507.1127K} presented an interesting analysis of the ellipticity in NGC~5139. The study illustrated that the cluster is more spherical at its core, while its ellipticity increases with the distance from the center.

Cluster rotation is also common. It was first detected in NGC~5139 ($\omega$~Centauri) and NGC~104 (47 Tucanae) by \citet{1986A&A...166..122M} using radial velocity (RV) measurements. The respective rotation velocities are $v_{\mathrm{max}} = 8 \, \mathrm{km \,s^{-1}}$ and $6.5 \, \mathrm{km \,s^{-1}}$. Further investigations by \citet{2017ApJ...844..167B} and \citet{2019MNRAS.485.1460S} revealed that the rotation axes of both clusters are inclined with respect to the line of sight at angles of $39.2 \pm 4.4^{\circ}$ and $33.6 \pm 1.8^{\circ}$, respectively.

A more recent analysis for 25 GCs carried out by \citet[hereafter: M23]{2023A&A...671A.106M} using 500,000 spectra of 200,000 stars obtained with the Multi Unit Spectroscopic Explorer (MUSE) \citep[henceforth: K18]{2018MNRAS.473.5591K}, revealed that at least 21 of them rotate. It is important to note that the different studies analyzing rotation did not necessarily analyze the same angular scales. As an example, the angular scales analyzed in this study are larger by 4 to 18 times than those investigated by M23. This distinction is crucial because the orientation of the rotation can vary depending on the specific scale that is studied. For NGC~5139, this is so extreme that its center rotates in the direction opposite to the global rotation \citep{2024MNRAS.528.4941P}. 

Considering that the vast majority of GCs are not spherical and that large-scale rotation has been detected in most of them, it is natural to ask whether these two phenomena are correlated. The first observational evidence suggesting a correlation between these two variables was found by \citet[hereafter: F14]{2014ApJ...787L..26F}. By comparing RV maps obtained with the VIRUS-W\footnote{Wide field visual integral-replicable unit spectrograph } Spectrograph \citep{2012SPIE.8446E..5KF} with the ellipticities of 11 GCs observed using the ACS/WFC camera of the \textit{Hubble} Space Telescope (\hst), F14 concluded that the geometry of clusters is primarily determined by their rotation. These findings were confirmed by K18, who concluded that this correlation indeed exists, based on a larger sample of GCs and RV measurements. However, \citet[henceforth: S19]{2019MNRAS.485.1460S} concluded that there is no significant evidence of a relation between the geometry and the rotation of GCs. This conclusion was derived based on a compilation of 45,561 RVs of individual stars from a sample of 109 GCs, including data from \gaia\ DR2 and multiple ground-based spectrographs, along with the ellipticities measured by C10. The main difference between these studies is that both F14 and K18 measured the ellipticities of the GCs using \hst\ images, while S19 used ellipticity measurements previously obtained by C10 based on the 2MASS catalog \citep{2006AJ....131.1163S}.  As we show below, it is likely that this difference contributes to the observed difference among their results.

As the distance increases, the precision in the measurement of the cluster ellipticities (and position angles) decreases because it becomes increasingly difficult to detect the cluster members and observe stars near the cluster center. C10 stated that the resolution of 2MASS (1 arcsec per pixel) only allowed them to explore the outer region of each cluster. This limitation similarly affects the measurements of \citet{1987ApJ...317..246W}, as the data were obtained using photographic plates in a 1.2 m Oschin Schmidt telescope.  Among the telescopes used for measuring the ellipticities of GCs, the \hst\ stands out with the highest resolution, with $\sim 0.05$ arcseconds per pixel in the F606W and F814W filters. 

In this context, it might be useful to employ \gaia\ \citep{2016A&A...595A...1G} to measure the ellipticities for GCs to reanalyze the correlation between the orientation of the ellipses and rotation.  \gaia\ requires a separation of 0.23 arsec to resolve two equal-brightness sources in the along-scan direction \citep{2015A&A...576A..74D}. Even though its resolution is lower than that of the \hst, \gaia\ presents distinct advantages. It provides parallaxes and proper motions for a larger number of sources, which enables us to distinguish cluster members from field stars. This is particularly useful for clusters located in regions with a high star density and for distinguishing the shape of the clusters at their borders, where field star contamination is high. The cluster FSR~1758 serves as a prime example because it is located within the Galactic bulge \citep{2019ApJ...870L..24B,2020A&A...635A.125Y}. Its position makes it difficult to visually distinguish the stars that belong to the cluster, and therefore, the traditional methods used to determine its geometry are not accurate.

This paper provides a detailed analysis of the ellipticities that were derived and used in \citet[hereafter: Paper I]{2024arXiv240208843C} to determine the membership of RR Lyrae stars in Galactic GCs. These measurements made use of the \gaia\ astrometry to assess the ellipticities of most of the GCs detected within the Milky Way. Our paper is structured as follows. Section \ref{sec:data} describes the data and the method we used to measure the ellipticities.  Section \ref{sec:results} provides an analysis of the ellipticities and a comparison with previous results from the literature. The relation between RVs and the geometry of GCs is discussed in Sect. \ref{sec:radial}. Section \ref{sec:Gaia} discusses the potential of \gaia\ for measuring the rotation of GCs, and Sect. \ref{sec:summary} provides a summary of the paper. Appendix \ref{sec:parameters} compiles the ellipticities and position angles from the literature.

\section{Description of the data and method}\label{sec:data}
The sample of clusters and their members used in this study for our analysis is exactly the same as was used in \citetalias{2024arXiv240208843C} to measure the ellipticity of GCs and to identify their population of RR Lyrae stars. This dataset was compiled by 
\citet[henceforth: VB21]{2021MNRAS.505.5978V}, who analyzed the astrometric parameters (parallax, positions, and proper motions) of all sources located in the vicinity of 170 galactic clusters\footnote{The sample of cluster members is available on 
 \href{https://zenodo.org/records/4891252}{Zenodo}.  } previously identified in the literature using \gaia\ EDR3.  As explained in \citetalias{2024arXiv240208843C}, we updated the VB21 dataset by cross-matching it with \gaia\ DR3 \citep{2022arXiv220800211G}. This update had no impact on the astrometry because \gaia\ DR3 and EDR3 share the same astrometric parameters. However, DR3 includes more sources with measured RVs.

Since the publication of the VB21 dataset, the \gaia\ collaboration has provided additional observations for NGC~5139 \citep{2023A&A...680A..35G}, acquired using the Sky Mapper instrument. This instrument allows us to capture images of the cluster, which enables a detailed exploration of its center and the observation of approximately half a million sources that were previously undetectable using the \gaia standard observation mode. The majority of sources observed with the Sky Mapper are associated with NGC~5139, but some of them are field stars.  The  area examined with the Sky Mapper spans $0.81$ degrees in radius, while the stars in the VB21 sample, with probabilities higher than $50\%$, only extends up to $0.67$ degrees. Distinguishing between cluster members and field stars in the sample provided by the Sky Mapper would require a membership analysis, which is beyond the scope of this paper. As a result, we did not include these stars in the measurement of the ellipticity of NGC~5139.

The procedure we used to measure the ellipticities of the GCs was described in Section 3.1 of \citetalias{2024arXiv240208843C}. We therefore explain the method only in general terms here and focus on comparing our results with prior research in the literature.  The approach used to measure the ellipticities involves modeling the 2D geometry in right ascension and the declination of each GC using a principal component analysis (PCA). The PCA determines the length and orientation of two orthogonal vectors on the sky plane, and we used these two vectors to define an ellipse.  We denote the major axis of the ellipse as $a$ and the minor axis as $b$. The eigenvectors ($\nu_{a},\nu_{b}$) correspond to the axes of the ellipse, and the eigenvalues ($\lambda_{a},\lambda_{b}$) provide a measure of their length.  The position angle was measured counterclockwise from the x-axis and is equal to $\theta = \arctan(\nu_{a,y}/\nu_{a,x})$, and the ellipticity is $\epsilon = 1 - \sqrt{\lambda_{b}/\lambda_{a}} = 1 - b/a$.  To determine the mean value and uncertainties of the parameters, we bootstrapped the sample of cluster members 1000 times. The ellipticity and position angle were calculated as the mean value of the distribution resulting from the bootstrap, and the uncertainties correspond to the standard deviation. Due to projection effects, our measurements set a lower limit to the true ellipticity. For example, if the major axis of an ellipse is aligned with the line of sight, the measured ellipticity corresponds to that of a circle.

\section{Results}\label{sec:results}
\begin{figure*}[h]
    \centering
    \includegraphics[scale=0.32]{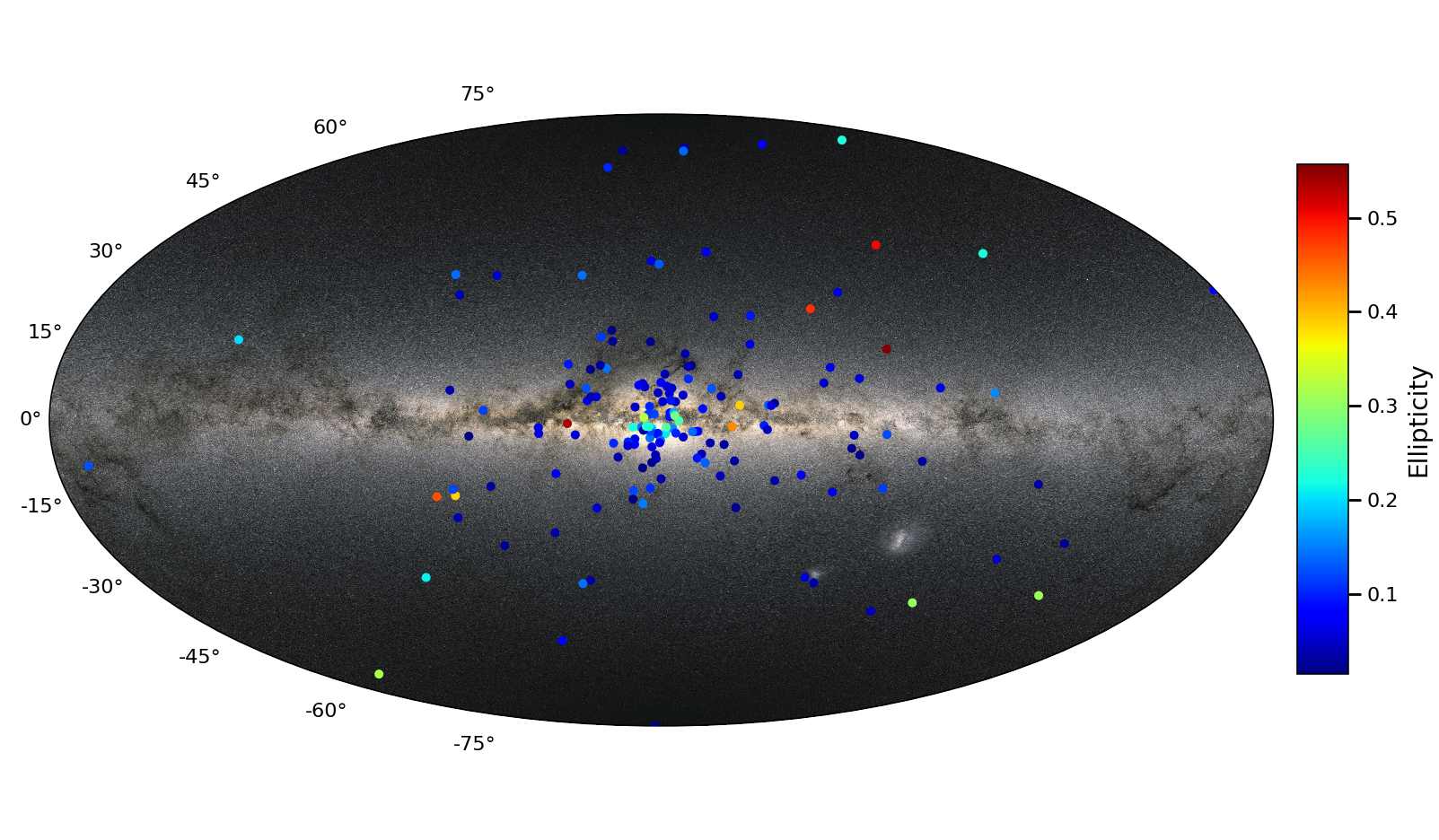}
    \vspace*{-5mm}
    \caption{Sky distribution in Galactic coordinates of all the GCs in our sample. Each cluster is color-coded with the ellipticity derived in this paper. The plot uses the Aitoff projection. The background image represents the 1.8 billion stars observed by \gaia\ and was generated by ESA/Gaia/DPAC.} 
    \label{fig:mw_ellipticty}
\end{figure*}

Using the method presented in the previous section, we were able to measure the ellipticity of 163 GCs.   Figure \ref{fig:mw_ellipticty} displays the distribution of cluster ellipticities in the sky.  Based on \gaia\ astrometry, we measured the ellipticities of FSR~1758 and BH~140 for the first time with a significance (defined as $\epsilon/\sigma_\epsilon$) greater than three and with more than 1000 detected members for both of them. Previously, the ellipticity of these clusters could not be measured because of the high density of stars surrounding them. BH 140 is located near the Galactic plane \citep{2018A&A...618A..93C}, and FSR 1758 resides in the Galactic bulge \citep{2020A&A...635A.125Y}.

\begin{figure}[h]
    \centering
    \includegraphics[scale = 0.75]{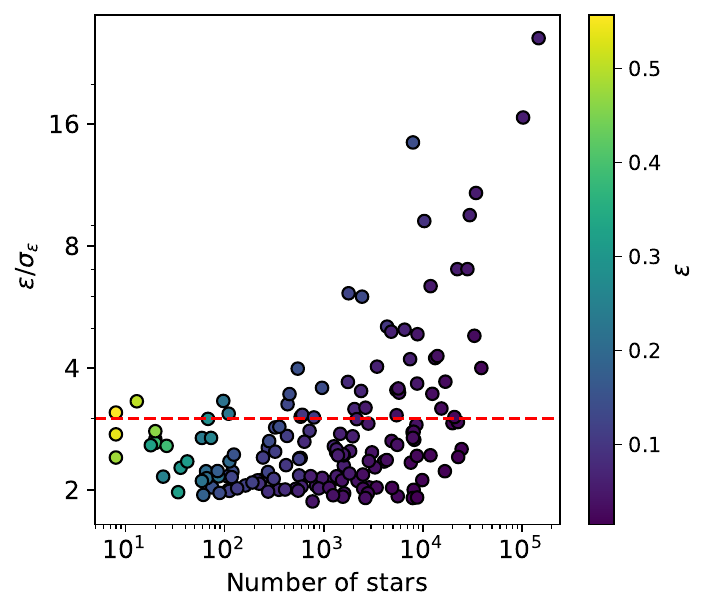}
    \caption{Ellipticity over the error as a function of the number of stars used to compute them. The red line indicates $\epsilon/\sigma_\epsilon = 3$. All points are colored according to their ellipticity values.}\label{fig:statistics_ellipses} 
\end{figure}

Figure \ref{fig:statistics_ellipses} illustrates the relation between the number of stars and the significance of the ellipticity for all clusters. As anticipated, the significance ratio increases with the number of cluster members.  The median number of stars used to determine the ellipticities per cluster is 1393, but it can reach 147,516 for clusters such as NGC~5139 and be as low as 8 for Ryu~879. The specific number of stars used for each cluster can be found in Table \ref{tab:ellipticities}.  Out of the 163 clusters for which we measured the ellipticity, only 46 clusters have $\epsilon/\sigma_\epsilon > 3$. We caution that clusters with a small number of members and a high eccentricity may require a further detailed inspection. Two of the 22 clusters with fewer than 100 stars and $\epsilon/\sigma_\epsilon > 2$ have previously published ellipticity measurements: UKS~1 and Terzan~12. These measurements were taken by C10. Our results for these clusters are consistent within $1\sigma$, although this might be attributed to the large uncertainties in the C10 data.  Additionally, Fig.\,\ref{fig:statistics_ellipses} suggests that clusters with particularly few stars have a particularly high $\epsilon$. 

\begin{figure}[h]
    \centering
    \includegraphics[width=1\linewidth]{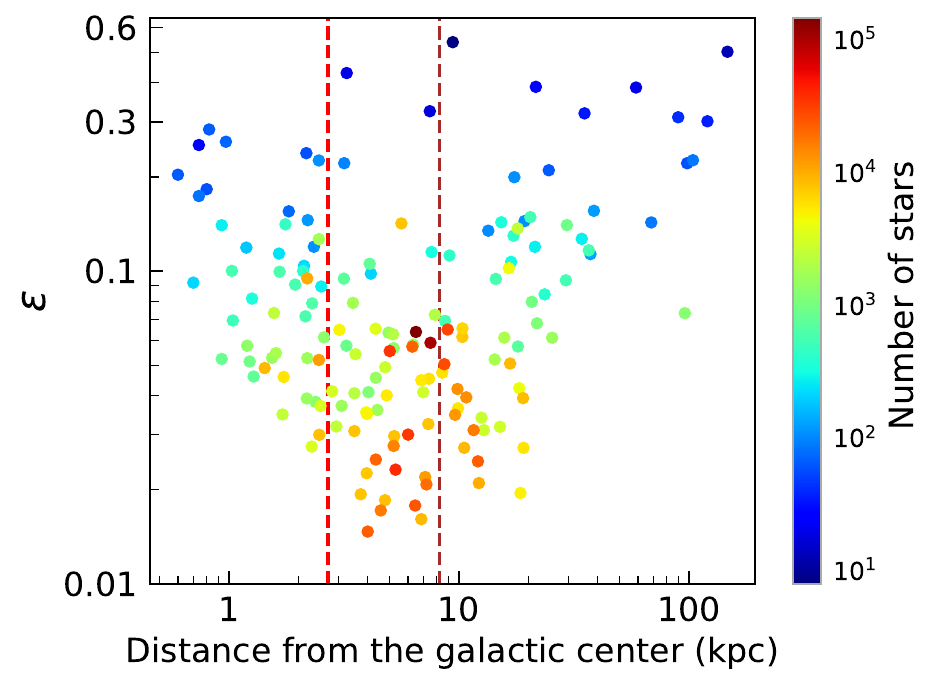}
    \caption{Ellipticities as a function of galactocentric distance. Each cluster is color-coded according to the number of stars used to determine its ellipticity. The red line marks a galactocentric distance of 2.7 kpc. Clusters with distances below this limit were previously considered members of the Galactic bulge by \cite{2010ApJ...721.1790C}. The brown line represents the galactocentric distance of the Sun \citep{2023AstL...49..493B}.} 
    \label{fig:distances_gc}
\end{figure}

Figure \ref{fig:distances_gc} shows the ellipticities of 157 out of 163 clusters in our dataset, plotted against their galactocentric distance ($R_{G}$). The distances were obtained from the Galactic GC Database Version 4\footnote{\url{https://people.smp.uq.edu.au/HolgerBaumgardt/globular/}}. The figure shows a clear correlation between the number of stars that are detected in a cluster and the ellipticity that is determined. At the same time, the number of detected stars is highest near $R_{G} \approx 8$ kpc, that is, around the Sun's galactocentric distance, likely because member stars are more readily resolved in GCs close to the Sun while also featuring smaller astrometric uncertainties. This also explains the trend toward fewer member stars in GCs located at galactocentric distances below $\sim 4$ and above $\sim 12$ kpc, which leads to a cone-shaped distribution. We therefore consider the trend toward higher ellipticities at short galactocentric radii spurious and caused by observational selection effects, and we cannot confirm the claim by \citet{2010ApJ...721.1790C} that clusters are more elliptical near the bulge. Our results suggest that a minimum number of stars of about $10^{3}$ is required to measure ellipticities faithfully. 

A particularly interesting cluster is NGC~6544, for which we find an ellipticity of $0.142 \pm 0.010$. It is located at a distance of $R_G = 5.62 \pm 0.06$ kpc. Clusters with a similar number of stars ($10^{3}$ - $10^{4}$) and with $R_{G} < 10$ kpc typically have a much lower ellipticity, with an average of $\epsilon = 0.09$.

In high-density regions where field stars are not easily distinguishable from cluster members, it is preferable to use \gaia\ data to assess the geometry of GCs. However, for clusters in which \gaia\ detects only a limited number of stars, traditional methods might offer more suitable alternatives for measuring their ellipticities.

\begin{figure*}
    \centering
    \includegraphics[scale=0.32]{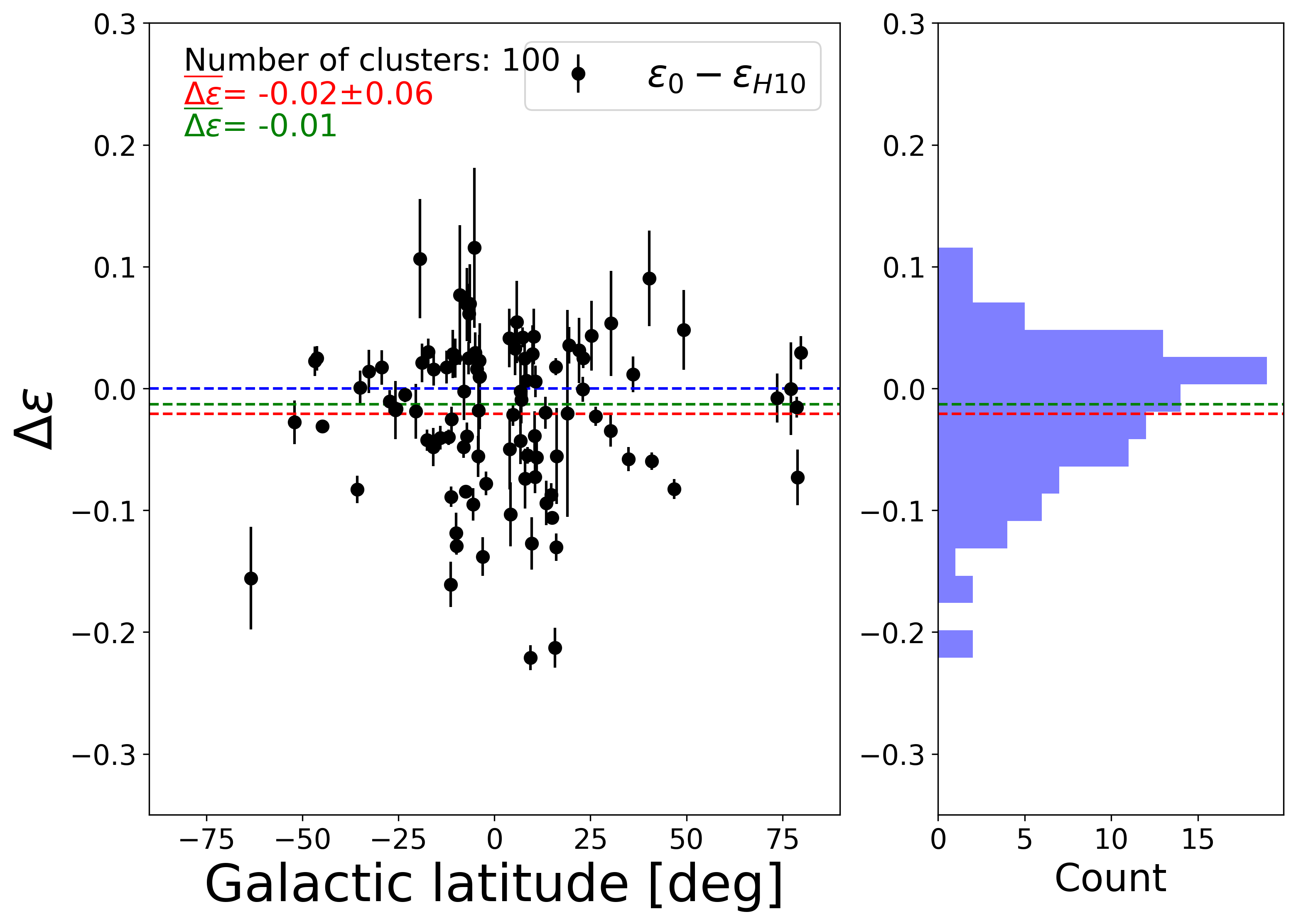}
    \includegraphics[scale=0.32]{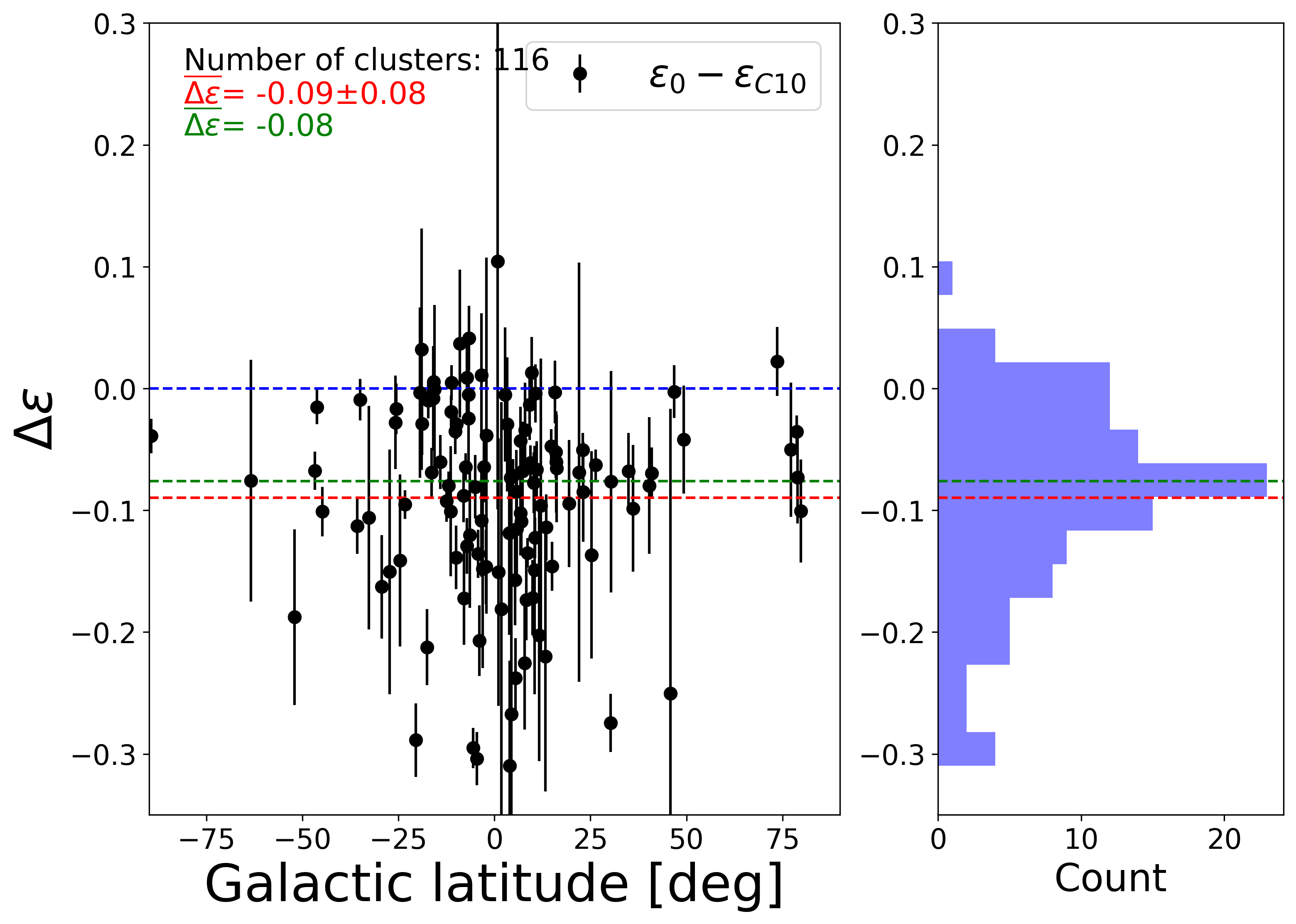}
    \includegraphics[scale=0.32]{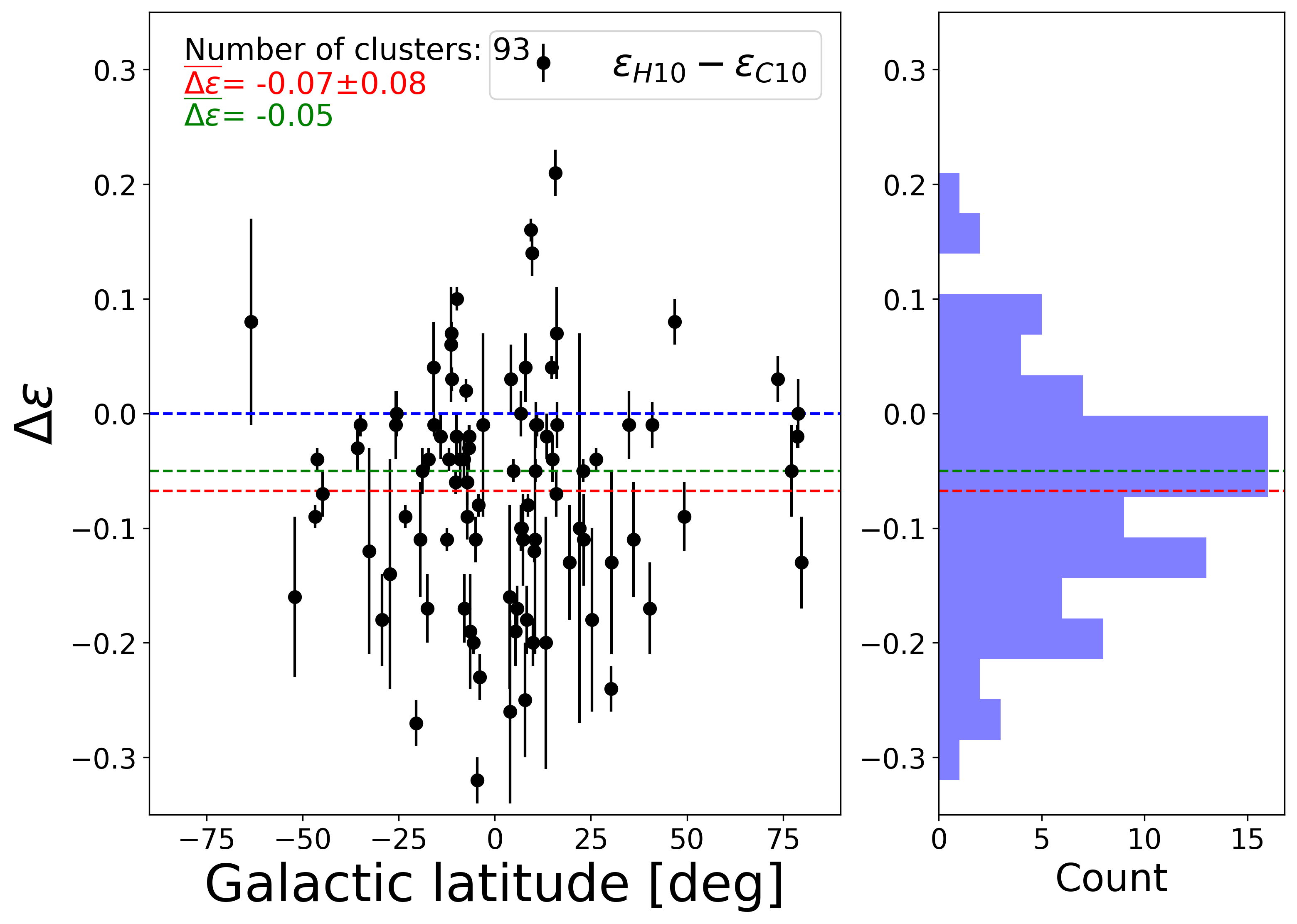}
    \caption{Comparison of the ellipticities obtained in different studies. All clusters present in both datasets are included, regardless of the number of stars considered in the PCA analysis. The ellipticities measured in this study are denoted by $\epsilon_{0}$. The error bars represent the sum in quadrature of the uncertainties in both measurements, except for comparisons involving H10, as the catalog does not report uncertainties. As a reference point, we added a dashed blue line around zero. The dotted red line indicates the mean difference between the ellipticities, and the green line shows the median. The red numbers displayed in the top left corner of each panel indicate the mean difference and its standard deviation.  The green numbers represent the median difference.     A histogram is placed on the right side of each figure to facilitate visualization of  $\Delta \epsilon$.    }
    \label{fig:ellipticities}
\end{figure*}

Figure \ref{fig:ellipticities} provides a comparison with previously reported results in the literature. However, we note that this comparison is not straightforward because the ellipticity may change as a function of the angular distance from the cluster center and because the ellipticity values reported in different studies were measured at different angular scales. For example, according to \cite{2021MNRAS.507.1127K}, the ellipticity of NGC~5139 extends from $\epsilon \sim 0.02$ at 40 arcmin from the center, and increases to $\epsilon \sim 0.6$ at the tidal tails of the cluster, 240 arcmin apart from the center. Previously, \citet{1983A&A...125..359G} measured $\epsilon = 0.05 \pm   0.010$ at 28 arcmin and \citet{2020ApJ...891..167C}  $\epsilon = 0.045$ at 30 arcmin. Our measurement indicates $\epsilon = 0.064 \pm 0.002$ at 40.2 arcmin.

\begin{figure*}
    \centering
    \includegraphics[scale=0.25]{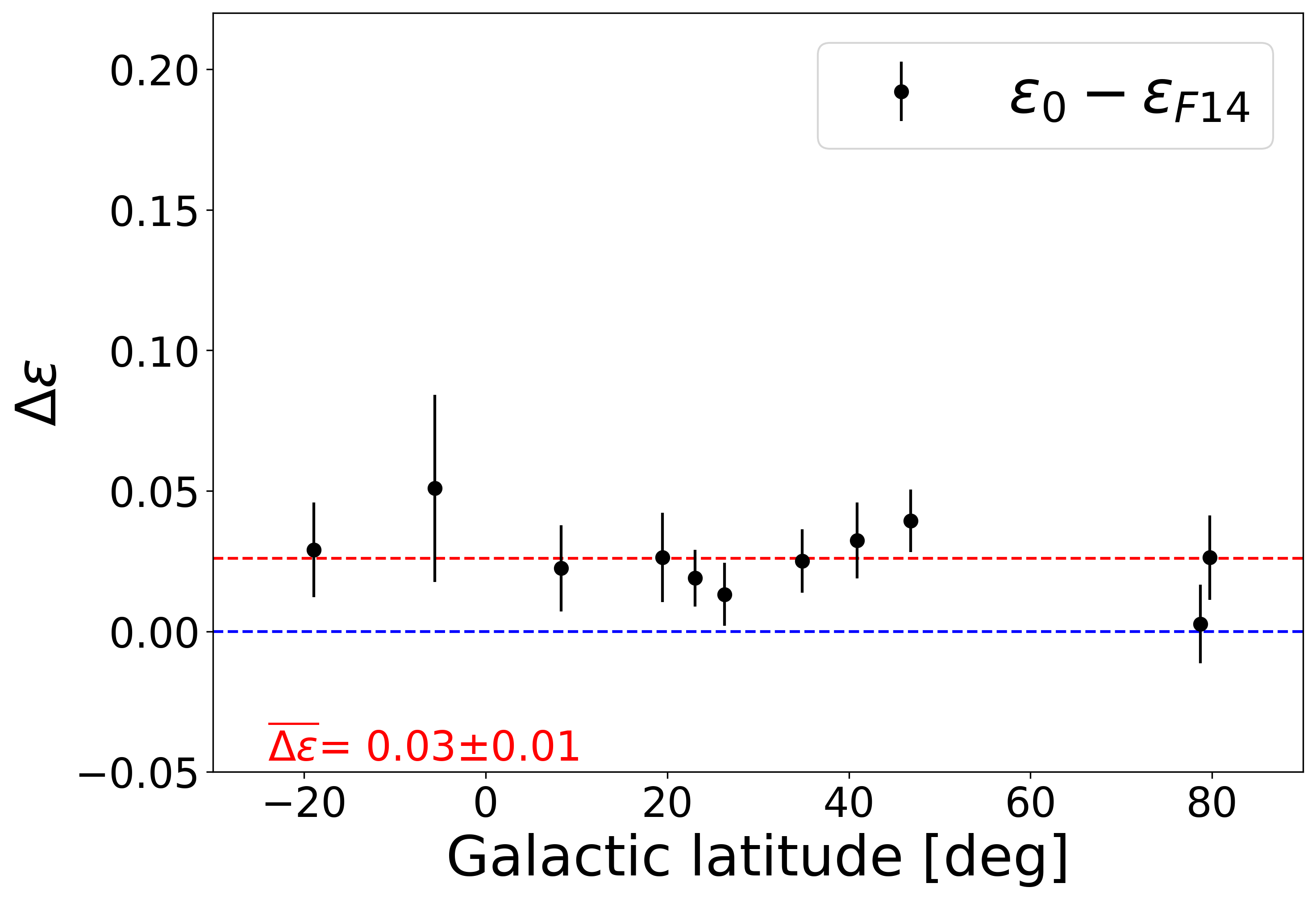}
    \includegraphics[scale=0.25]{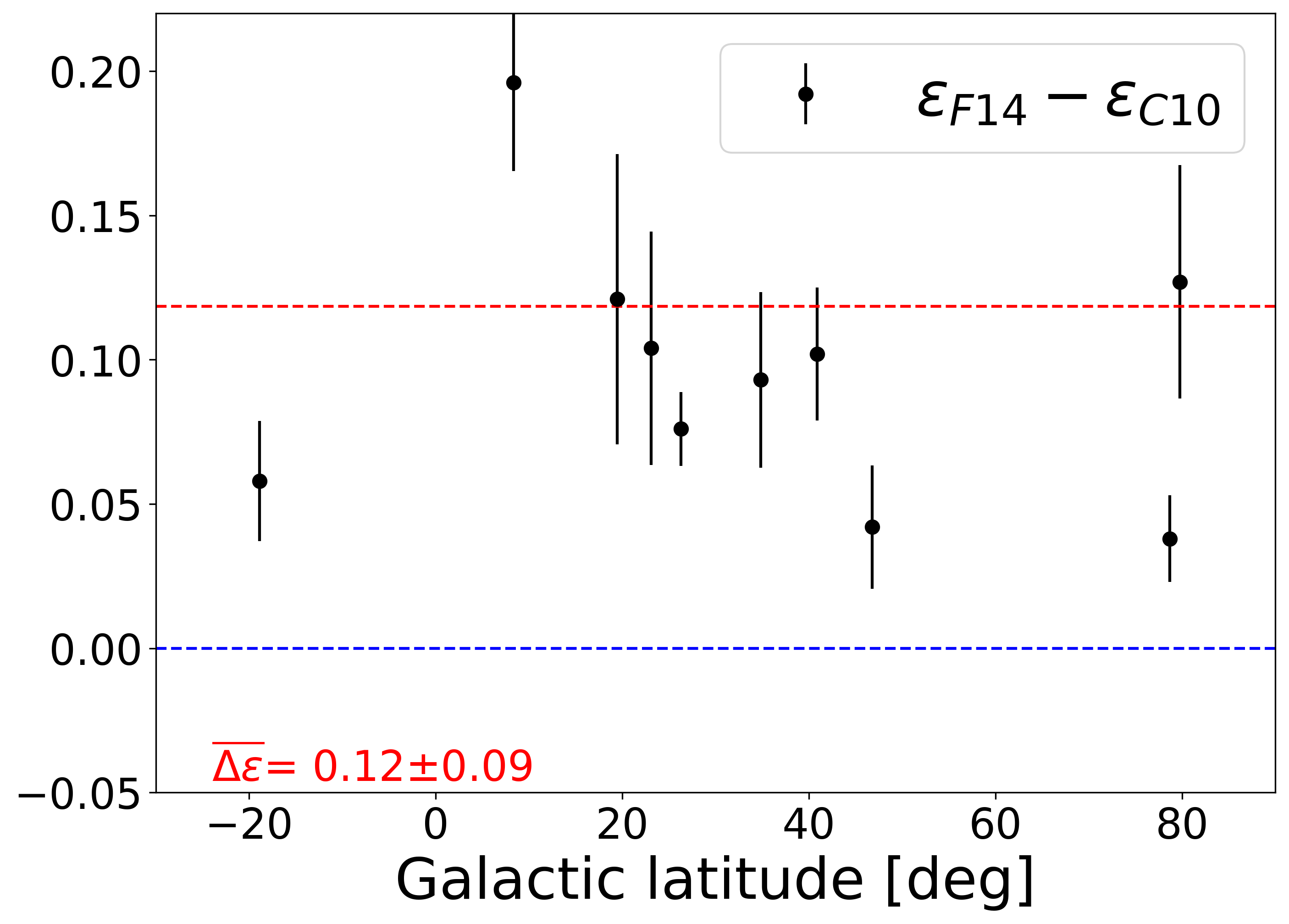}
    \includegraphics[scale=0.25]{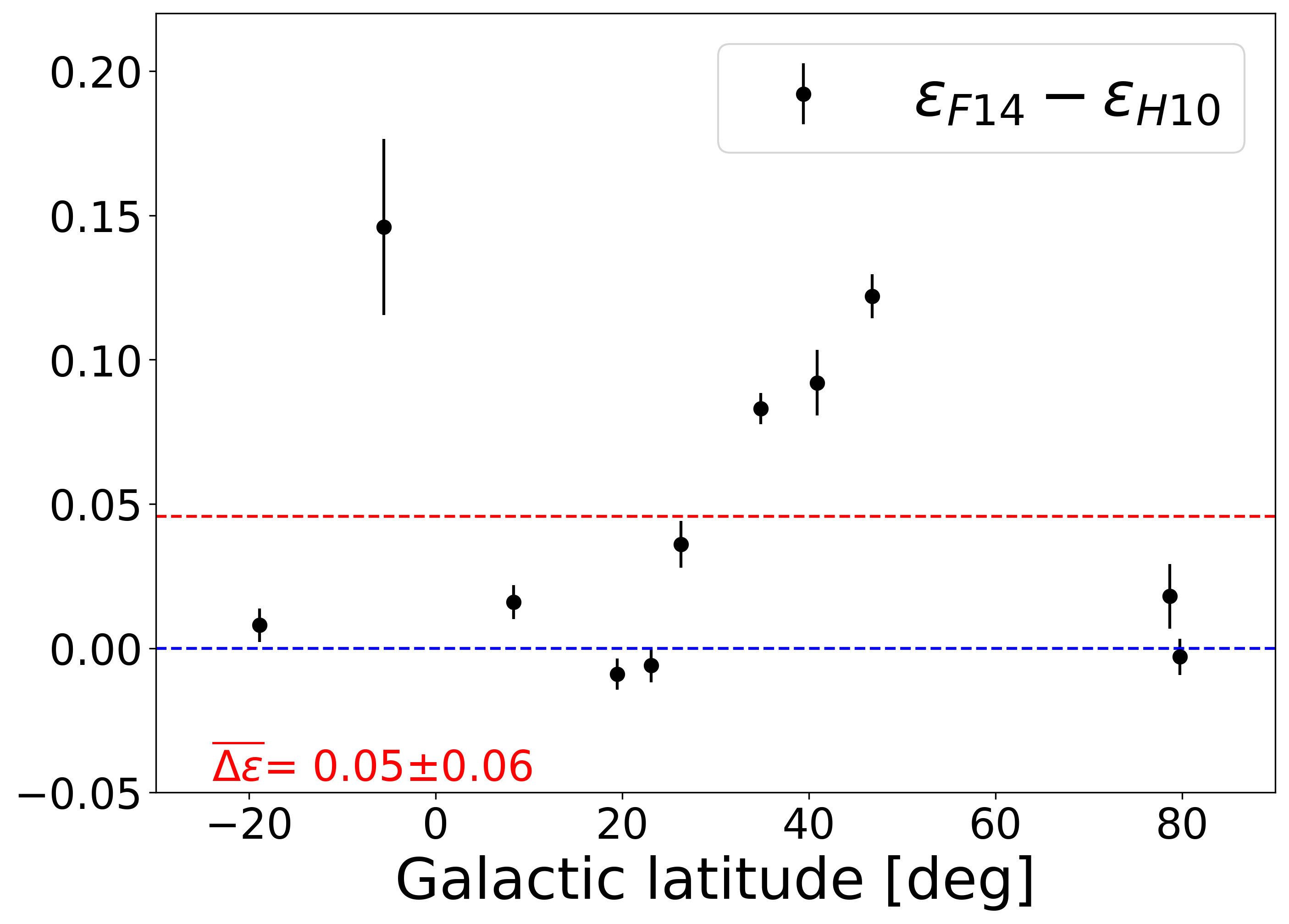}
    \caption{Same as in Fig. \ref{fig:ellipticities}, but comparing the ellipticities from this study, H10, and C10 with those obtained using \hst\ photometry. } 
    \label{fig:ellipticities_HST}
\end{figure*}

Figure \ref{fig:ellipticities_HST} shows the good agreement between the ellipticities of the 11 GCs studied by F14 using \hst\ imaging and our results based on \gaia, aside from a small mean difference of $\Delta \epsilon = 0.026 \pm 0.013$ (our results tend to be slightly more elliptical).  Fortunately, all  datasets contain the 11 clusters examined by F14 with the \hst, making it possible to clearly see the differences in the ellipticity values reported by each study.  Figure \ref{fig:ellipticities} shows that the comparison of our results with ground-based studies exhibits large scatter, likely due to the lower spatial resolution of the telescopes.

\begin{figure*}
    \centering
    \includegraphics[scale = 0.5]{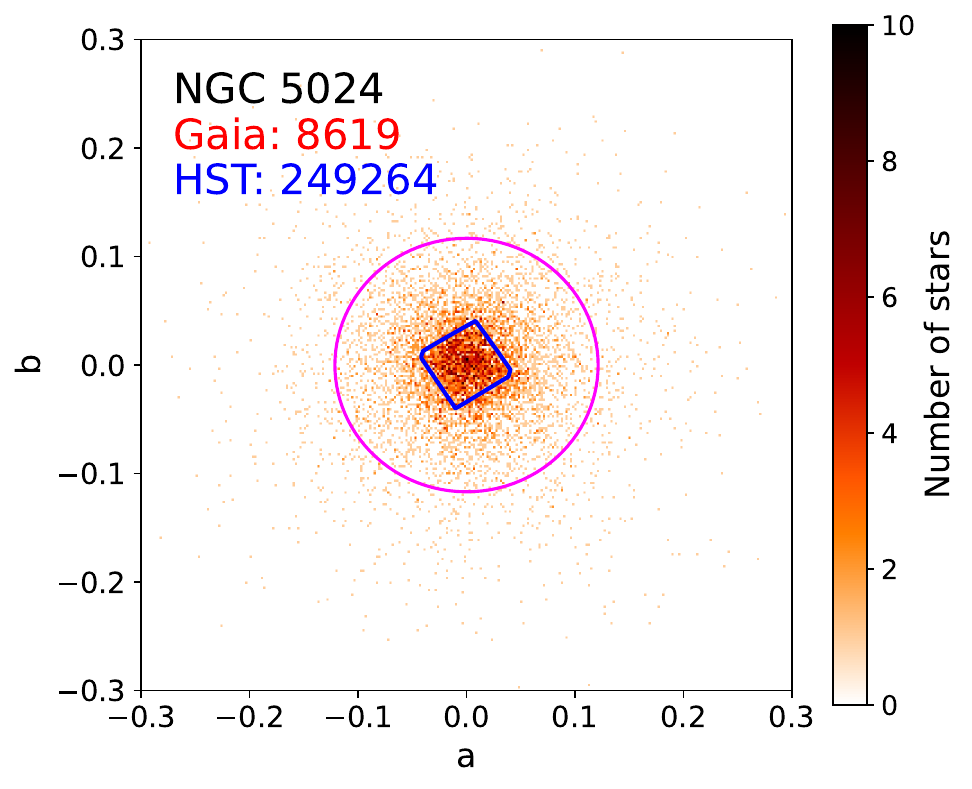}
        \includegraphics[scale = 0.5]{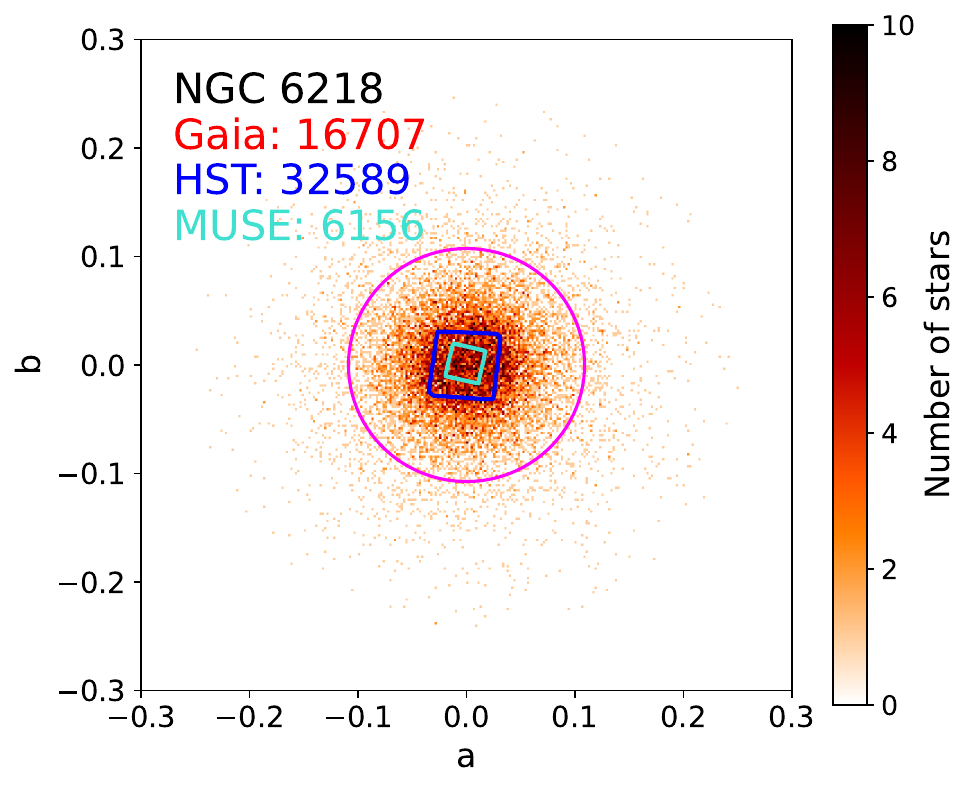}
    \caption{Comparison of the samples used to determine ellipticities in this study and that of F14. The color bar is used to indicate the density of sources in the \gaia\ data. The magenta ellipse represents the results of the PCA analysis. The major and minor axes are aligned with the plot axes, and the stars were rotated to match this alignment. The blue contour shows the data used by F14 to measure their ellipticities, which was originally compiled by \citet{2007AJ....133.1658S} using the \hst. The turquoise contour represents footprint of the stars in the M23 sample. The numbers at the top correspond to the number of stars in each sample.    }
    \label{fig:gaia_hst}
\end{figure*}

Figure \ref{fig:gaia_hst} shows a comparison between the sample of stars used to measure the geometry of NGC~5024 and NGC~6218 in this study and the sample provided by F14. For the clusters in common, the radius covered by \hst\ is about seven times smaller on average than that covered by \gaia, but the median number of stars in the \hst\ sample is nine times larger than that obtained with \gaia. This difference can be attributed to two main factors: completeness and resolution. The \gaia\ observations are subject to a complex selection function that is the subject of ongoing research (see \citet{2023A&A...669A..55C} for the example of NGC~1261). A dominant factor of the selection function is that when the \gaia\ observational capacity of 1,050,000 objects per square degree \citep{2016A&A...595A...1G} is exhausted, the brightest stars are prioritized. As a result, the \gaia\ completeness is lowest at the center of clusters and for sources with $G>18$. Additionally, the higher spatial resolution of the \hst\ better distinguishes sources in the very busy GC centers, resulting in much larger numbers of stars despite the significantly smaller footprint.

Unfortunately, it is not possible to make a comparison similar to that shown in Fig. \ref{fig:gaia_hst} for the sample of clusters in C10 and H10 because the data used in these papers are not available.

Based on \gaia, we measured the ellipticities of 163 GCs with a similar precision as \hst.  The median axial ratio ($b/a$) for H10, C10, F14, and our sample are shown in Table \ref{tab:median_ratios}. Individual ellipticities and position angles are presented in Tables \ref{tab:ellipticities} and \ref{tab:position_angles}.

\section{Radial velocities and their connection with the ellipticities}\label{sec:radial}
As mentioned in the introduction, F14 identified a connection between the rotation of GCs and their ellipticities, revealing that the position angles of the ellipses, derived from both RV measurements and \hst\ images, are generally aligned. To compare the position angles measured by F14 with ours, we first transformed them to the same reference system. This can be achieved by using the following equation:
\begin{equation}\label{eq:f14}
 \theta^{\mathrm{T}} = (90^{\circ} - \theta)\bmod 180^{\circ},   
\end{equation}
where $\theta^{\mathrm{T}}$ is the F14 angle in our coordinate system, and $\theta$ is the angle originally measured by F14. When comparing our angles ($\theta_{0}$) with the F14 angles, it is important to note that both are only defined from $0^{\circ}$ to $180^{\circ}$. This implies that the maximum difference that can exist between them is  $90^{\circ}$. Consequently, the difference between them is given by

\begin{equation}\label{eq:2}
    \Delta \theta  = \min   \left (  |\theta_{0} - \theta^{\mathrm{T}} |, 180 - |\theta_{0} - \theta^{\mathrm{T}}| \right  ).
\end{equation}

On the other hand, both S19 and M23 measured the position angle of the rotation axis for the GCs in their sample. Thus, if there is a relation between the geometry of GCs and rotation, our angles should be correlated with the angle that is perpendicular to those measured by S19 and M23. These angles are defined from $0^{\circ}$ to $360^{\circ}$, and when we compare them with ours, we therefore need to remove the orientation. In this way, we avoid overestimating the differences between them.  Additionally, we need to transform them to our reference system. For these two transformations, we used the following equation:
\begin{equation}\label{eq:S19_M23}
 \theta^{\mathrm{T}} =  \theta \bmod 180 ^{\circ},
 \end{equation}
where $\theta^{\mathrm{T}}$ is the angle that we need for the comparison, and $\theta$ represents the angles originally measured by S19 or M23. To estimate the difference between our angles and those measured by S19 and M23, we used Eq. \ref{eq:2}.

\begin{figure*}[h]
    \centering
    \includegraphics[scale=0.385]{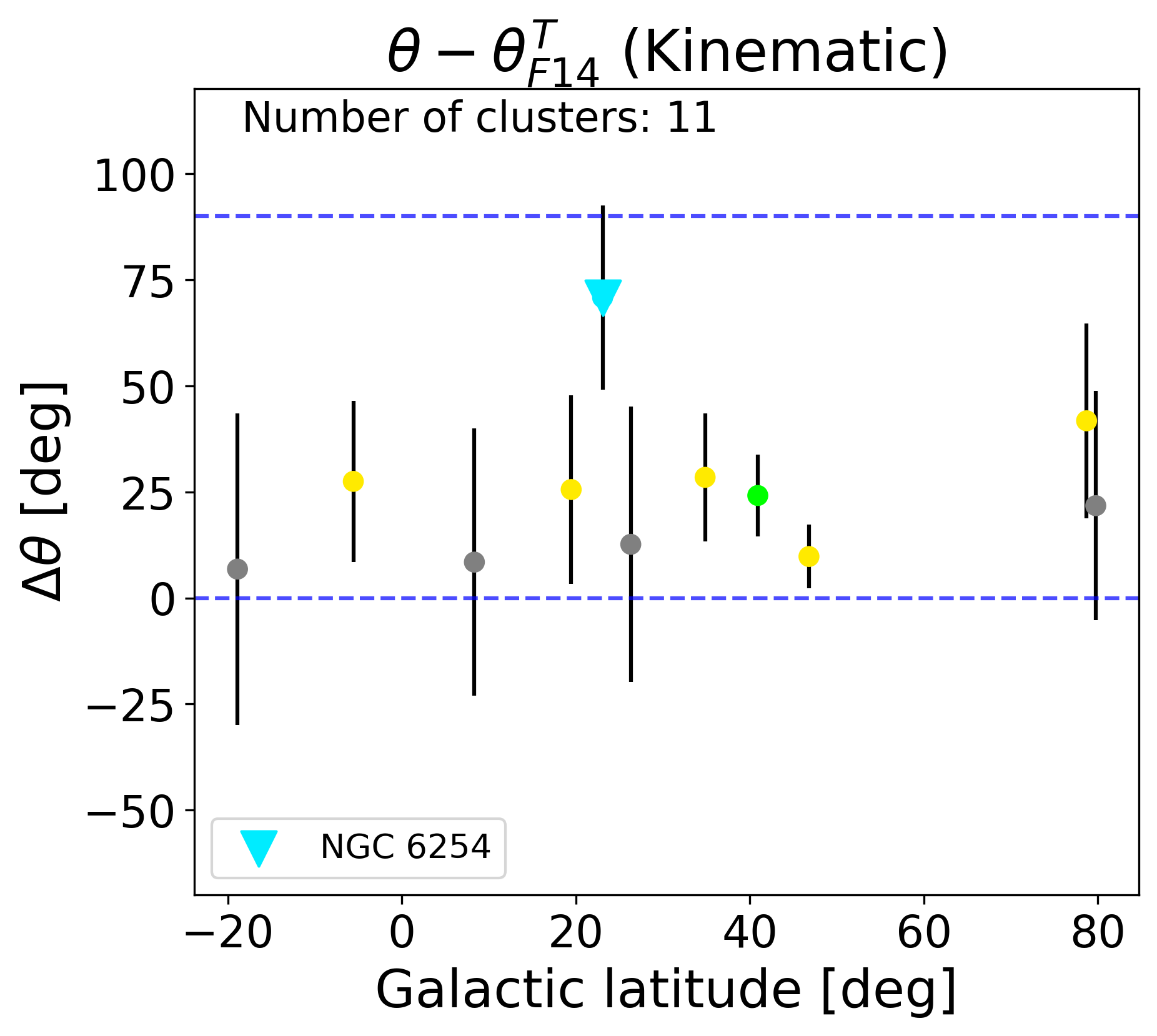}
    \includegraphics[scale=0.385]{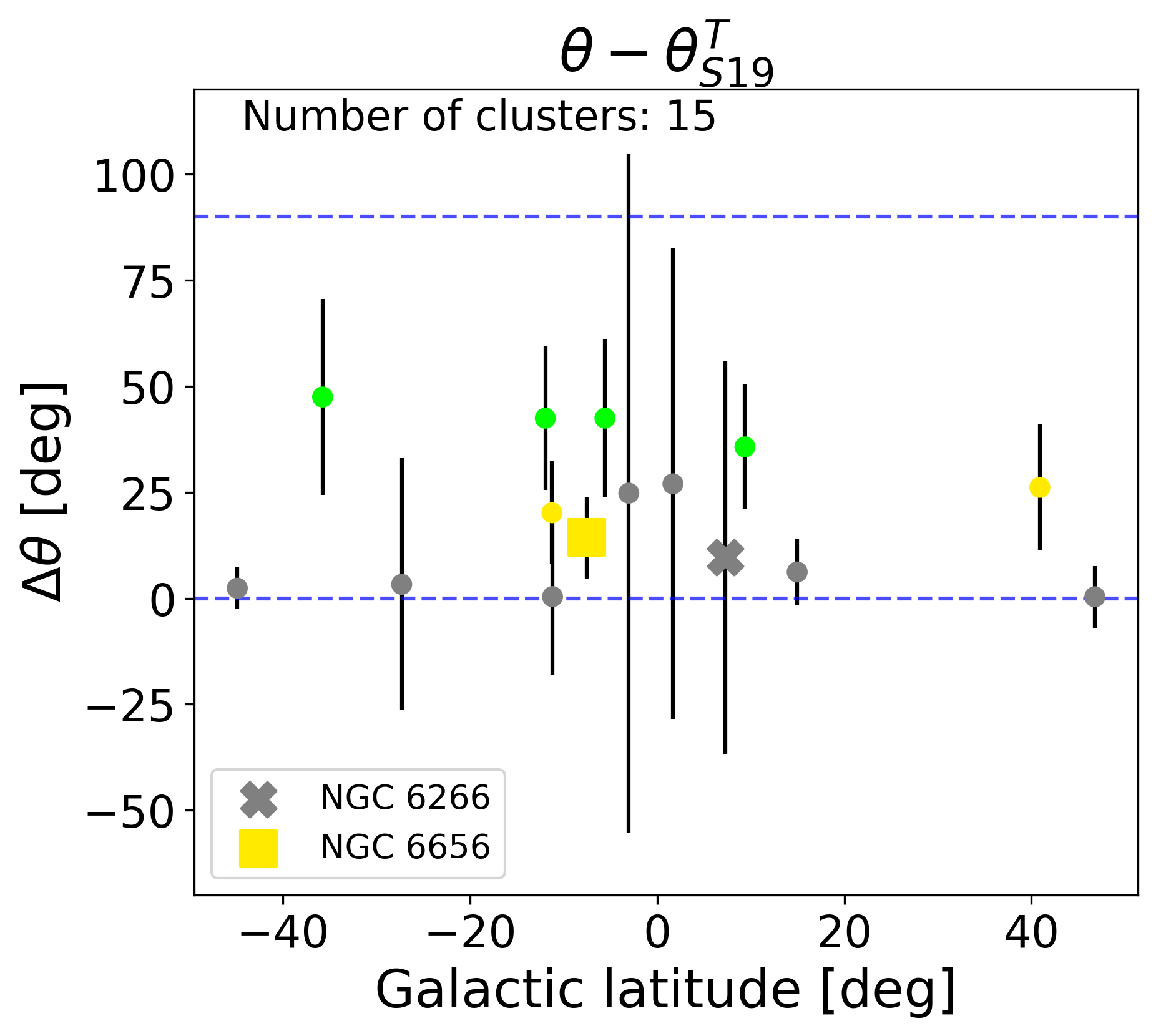}
     \includegraphics[scale=0.385]{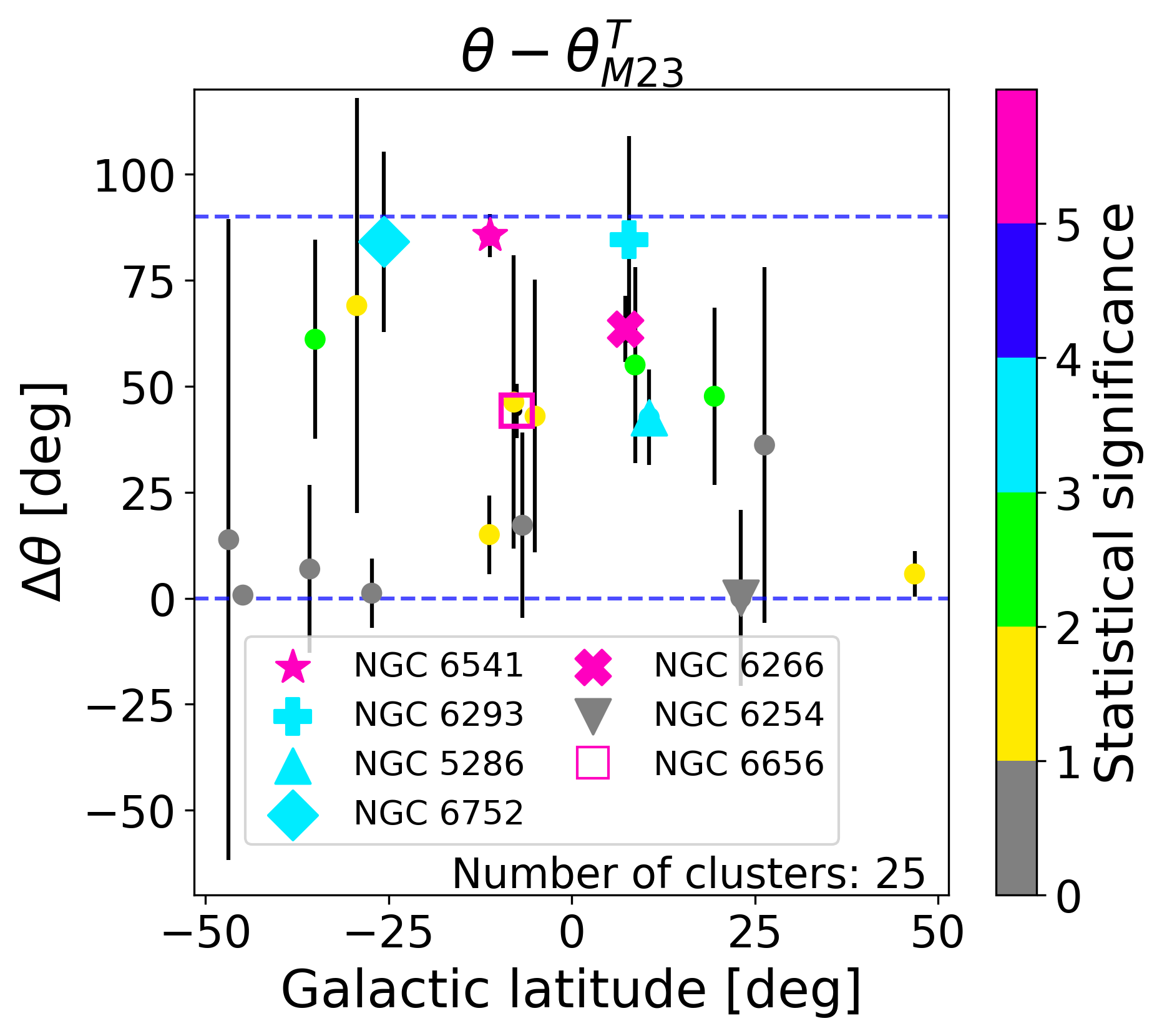}
    \caption{Comparison of the position angles measured in this study with those obtained from the literature. The color bar indicates the statistical significance of this difference, and the last bin contains all clusters for which this difference exceeds $5 \,\sigma$. Since our angles are not oriented, the maximum difference that can exist between our angles and those reported in the literature is $90^{\circ}$. The blue lines are placed as references at $0^{\circ}$ and $90^{\circ}$. Clusters for which the difference is greater than three standard deviations with respect to our angles are highlighted in all panels, even if this difference is with respect to only one dataset.  }
    \label{fig:angles_RVs}
\end{figure*}

\begin{figure}[h]
    \centering
     \includegraphics[scale=0.6]{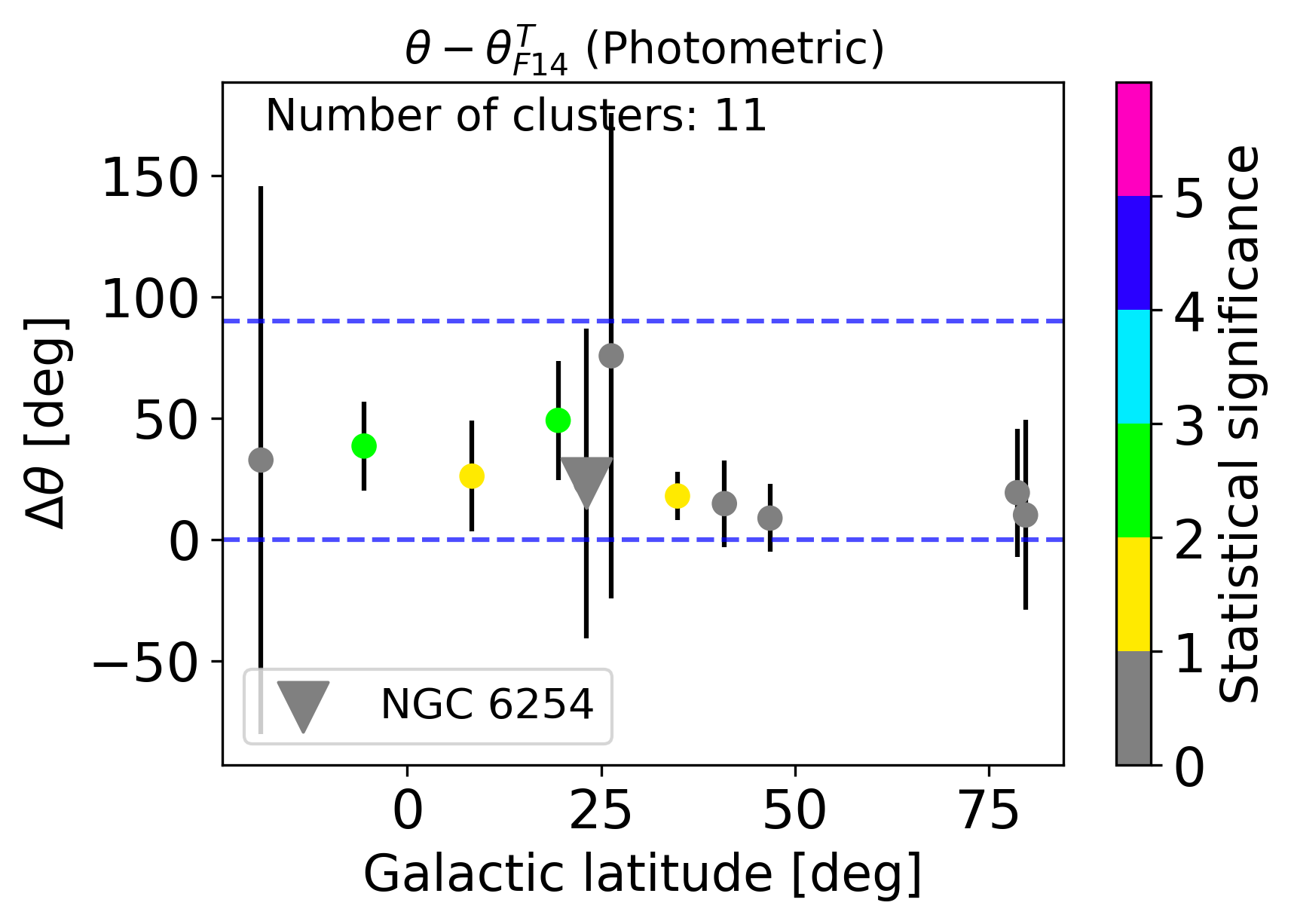}
    \caption{Comparison of the position angles measured in this study with those by F14 using \hst\ photometry. }
    \label{fig:angles_phot}
\end{figure}

Figures \ref{fig:angles_RVs} and \ref{fig:angles_phot} summarize the comparison between our position angles and the corresponding ones from the literature obtained using RV measurements and  \hst\ photometry. The data necessary to reproduce these plots can be found in Table \ref{tab:position_angles}. We defined the angle differences as statistically significant when they exceeded three standard deviations, with the standard deviation being the sum in quadrature of the uncertainties from both measurements. These clusters are highlighted in all panels, even when the difference is above our threshold with respect to only one dataset. Figure \ref{fig:angles_phot} shows that the position angles obtained with \gaia\ are consistent within $\sim 2\, \sigma$ with those measured by F14 using \hst\ photometry. By comparing our angles with the angles of F14 obtained using RVs,  we observed that for NGC~6254, this difference exceeds our threshold. However, this cluster does not stand out as an outlier when compared with the results of M23. None of the clusters in the S19 sample exhibit differences beyond our threshold.

Discrepancies in position angles may arise due to the use of different angular scales in the measurements. As shown by \citet{2018MNRAS.473.5591K} for NGC~7078, the position angle derived from RVs varies depending on the angular separation from the cluster center. The angular scales employed for measuring ellipticities in this paper are nine times larger on average than those used for M23 to measure rotation. In the M23 sample, six clusters show statistically significant differences: NGC~6266, NGC~6656, NGC~5286, NGC~6541, NGC~6293, and NGC~6752.  For the first two clusters, this difference disappears when using the angles from F14 or S19.  The remaining four GCs are not included in either the F14 or S19 datasets.  In particular, there is a difference of $42.64 \pm 9.69 ^{\circ}$ between our angle and the one reported by M23 for NGC~5286. For the clusters NGC~6541, NGC~6293, and NGC~6752, our angles are aligned perpendicularly to those reported by M23 within the uncertainties. It should be noted that the statistical confidence in the ellipticity of NGC~6293 is particularly low, being only $2\, \sigma$. We therefore considered that more evidence is necessary before we can conclude that the angles are perpendicular for this cluster. 

\begin{figure*}[h]
    \centering
    \includegraphics[scale = 0.5]{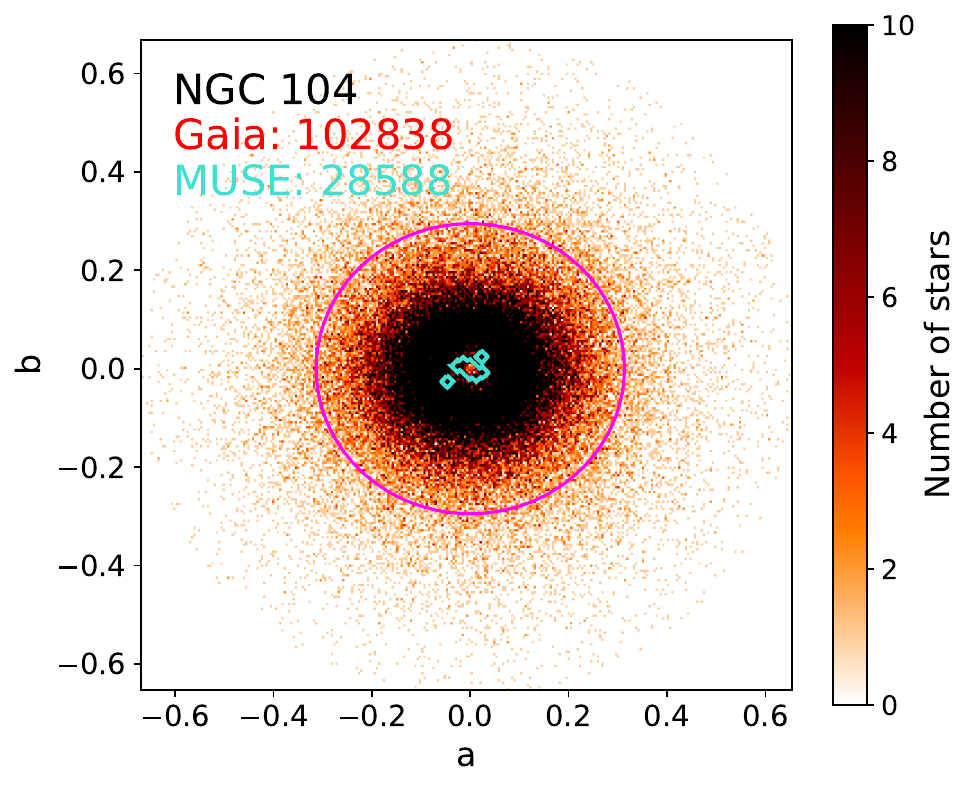}
        \includegraphics[scale = 0.5]{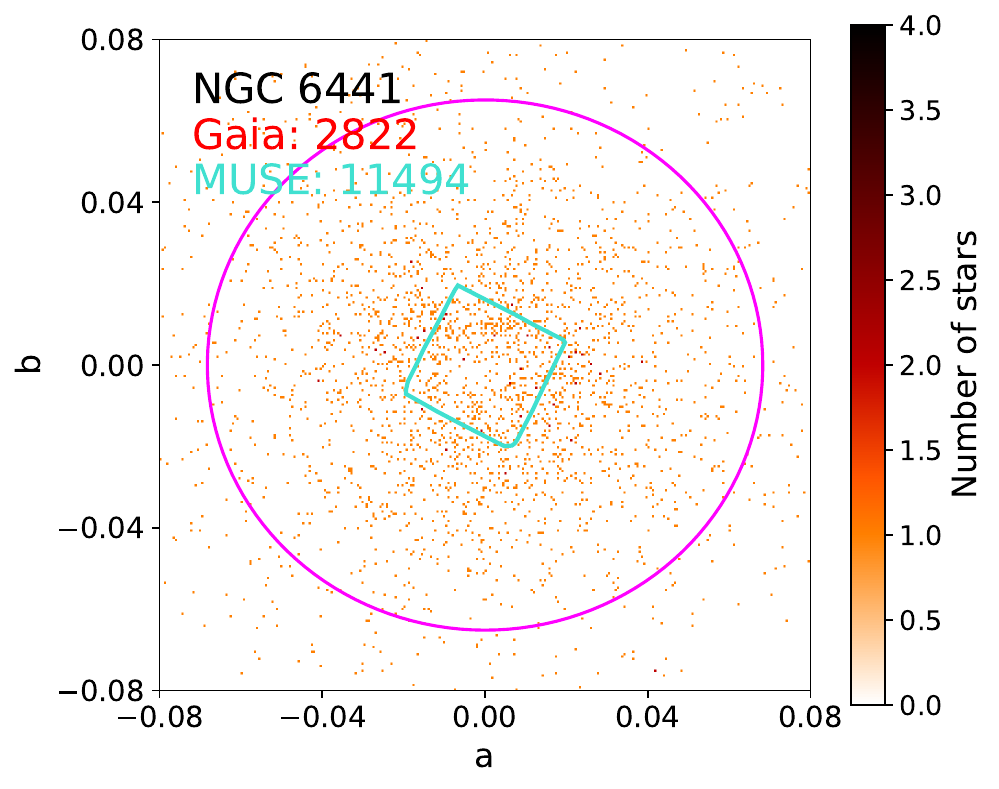}
    \caption{Comparison of the dataset used in this study to measure the position angles and the dataset used by M23. The color scheme is the same as in  Fig. \ref{fig:gaia_hst}.}
    \label{fig:gaia_muse}
\end{figure*}

Figure \ref{fig:gaia_muse} shows a comparison of the angular scales examined in this study to determine the position angles with those analyzed by M23 for two GCs. The data collected by M23 primarily target the central regions of clusters, while \gaia\ allows the analysis of entire clusters.  However, as shown in the figure, for NGC~104, the completeness of \gaia\ decreases near the region analyzed by M23 due to its limited angular resolution. For NGC~6441, the M23 observations are restricted to a radius that is smaller by about five times than that of \gaia, but the number of stars in the M23 sample is four times higher. 

The difference in the projection of the rotation axis and the position angles of our ellipses remains below three standard deviations for 100\%, 90\%, and 76\% of the clusters in the S19, F14, and M23 samples, respectively.  This suggest that there is a correlation between the geometric shape of GCs and their rotation. However, as shown previously, the  ellipticities and rotation were measured at different angular scales. As we show in the next section, there are three clusters for which we can measure rotation with \gaia\ at the same angular scale as the ellipticities. For these clusters, the projection of the rotation axis and the position angle derived from the PCA are aligned.

\section{Globular cluster rotation and \gaia}\label{sec:Gaia}
The RV spectrometer on board the \gaia\ satellite measured average RVs of 33,812,183 stars brighter than $G_{RVS} = 14$ mag across the full sky, including 10084 in GCs. The performance verification paper for RVs \citep{2023A&A...674A...5K} demonstrated that the RVs were abundant and precise enough to detect the rotation of NGC~104. This section analyzes the potential of \gaia\ to detect rotation in a larger sample of clusters. 

\begin{figure}[h]
    \centering
\includegraphics{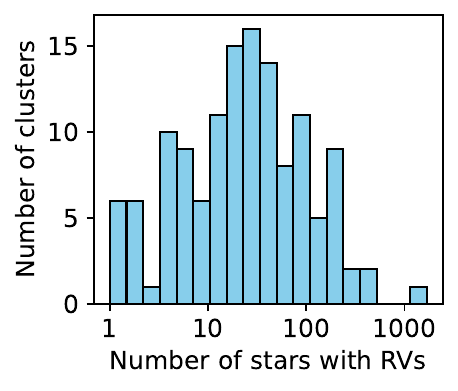}
    \caption{Number of RVs per cluster in \gaia\ DR3. }
    \label{fig:RV_numbers}
\end{figure}

We analyzed all clusters and cluster members with membership probabilities above 90\% in the VB21 sample. For 133 clusters, there is at least one RV measurement in \gaia\ DR3.  The distribution of the number of RV measurements in clusters can be found in Fig. \ref{fig:RV_numbers}. 

We divided the members of each cluster into two subsets using the minor axis of the ellipses as a reference and estimated the difference in the median RV among the two subsets.  We considered rotation to be detected when this difference exceeded three times the square-summed standard error on the median of the subsets. Following this procedure for all clusters in our sample, we detected the rotation of NGC~5139  and NGC~104 with a statistical confidence of seven and eight standard deviations.  By further dividing the sample of cluster members for NGC~5904 into different sections, we detected its rotation with a $3 \, \sigma$ confidence.  The absolute values of the rotation amplitudes for NGC~104, NGC~5139, and NGC~5904 are $5.01 \pm 0.61$, $6.65 \pm 0.97$, and $6.41\pm 1.95 \, \mathrm{km\,s^{-1}}$, respectively. These amplitudes are consistent within two standard deviations with those found by S19.

\begin{figure*}[h]
    \sidecaption
    \includegraphics[width=12cm]{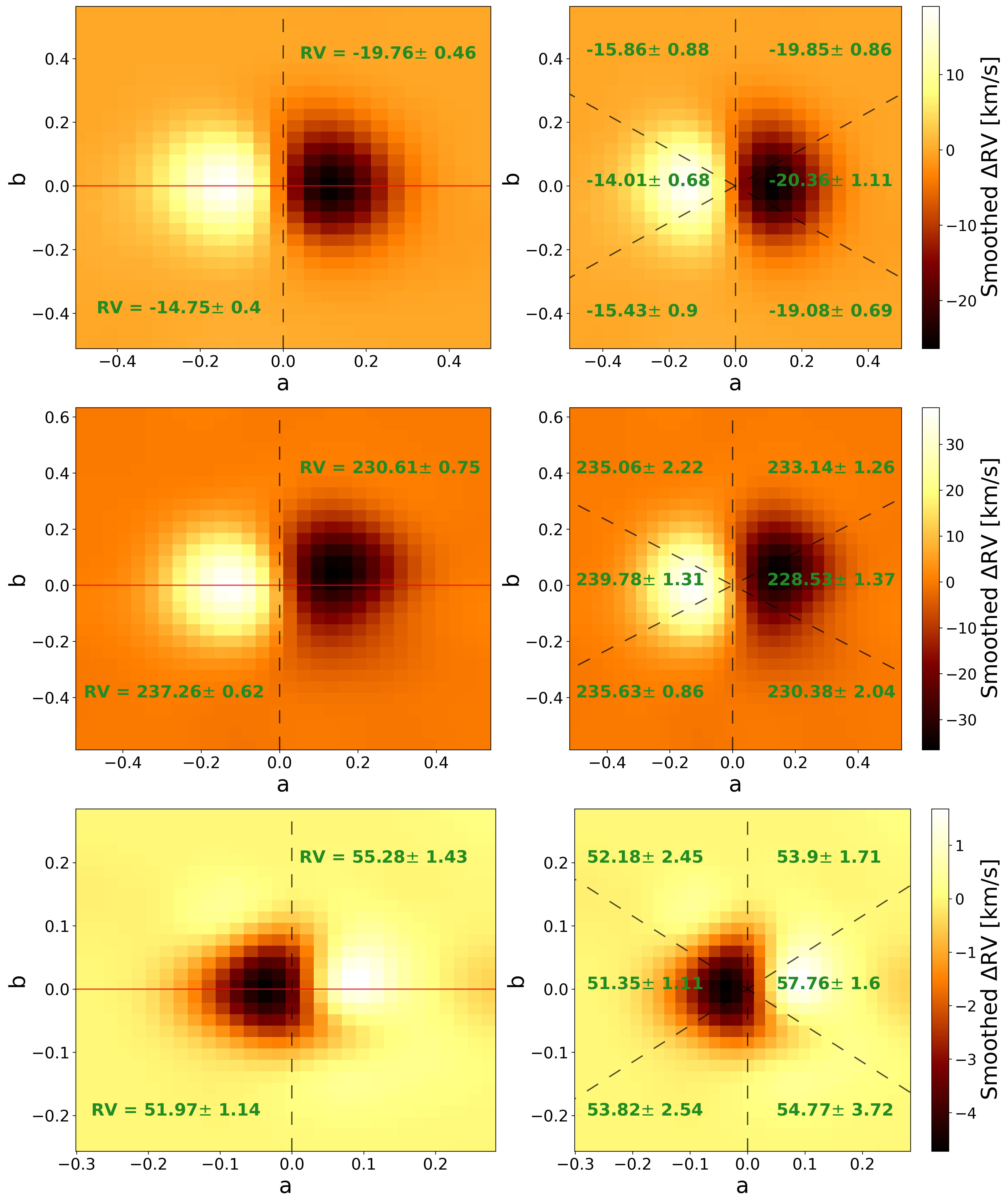}
         \caption{Radial velocity maps of NGC~104 (top), NGC~5139 (middle), and NGC~5904 (bottom). Left:  The red line is oriented in the direction of the major axis of the ellipse, and the dashed black line is aligned with the minor axis. The minor axis divides the sample of stars into two subsamples, with the median RV and the error on the median for each shown in green. The plot is color-coded based on the difference between the local value of RV and the median RV of the cluster. Right: Same set of stars as in the left panel. The cluster was divided into six subsections. Stars with an RV difference greater than $100 \, \mathrm{km\,s^{-1}}$ with respect to the median were not used to generate the RV maps. The plot is only used for a data visualization and not to measure the RV difference between the subsamples.  } 
   \label{fig:maps1}
\end{figure*}

The RV maps for these clusters are shown in Fig. \ref{fig:maps1}. The stars were rotated to align the orientation of the ellipse with the plot axes. If the measured ellipticity were caused by rotation, the minor axis would be expected to align with the projected rotation axis. The figure shows that they are  indeed aligned. 
\citet{2018MNRAS.473.5591K} concluded that for NGC~5139, the axes are perpendicular, although their study analyses the central region of the cluster. As mentioned in their paper, this difference may arise from the fact that the position angles derived from RVs change relative to the cluster center distance.

\begin{figure*}
\sidecaption
\includegraphics[width=12cm]{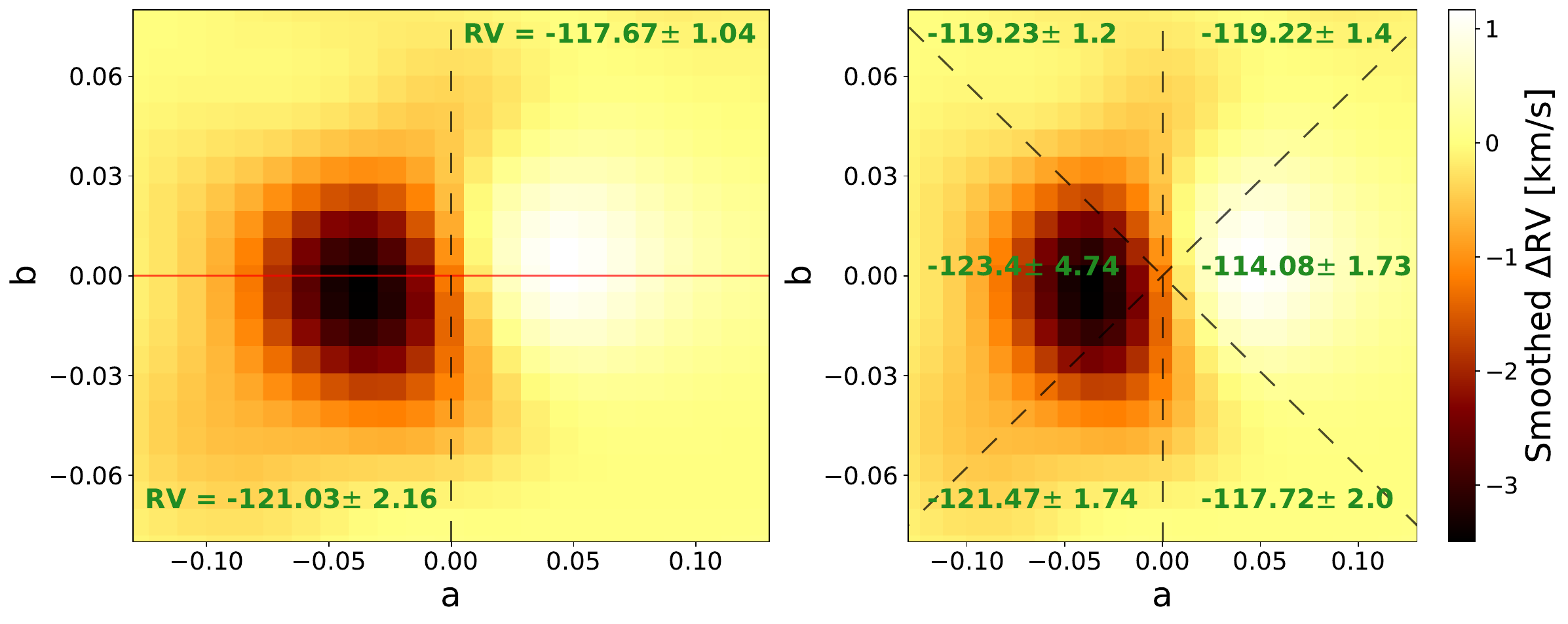}
    \caption{Radial velocity maps of NGC 6341. F14 concluded that this cluster rotates. In our case, the statistical uncertainties do not allow us to reach the same conclusion.  }
    \label{fig:NGC_6341}
\end{figure*}

We inspected the remaining 130 clusters individually to account for cases in which the axis of rotation is not aligned with the minor axis of the ellipses. We found no cluster for which this applied. However,  NGC~6341, for which F14 previously reported a detection of rotation, is an interesting case. Fig. \ref{fig:NGC_6341} shows that the RV map exhibits a pattern similar to the patterns found in the clusters for which we detected rotation. However, the statistical uncertainties on the RVs do not allow us to conclude that this cluster rotates. This could be related to the fact that there are only 103 stars with measured RVs for this cluster in \gaia\ DR3. 

The main reason why rotation has not been detected in more GCs is the low number of stars with RV measurements available in \gaia. The median number of RV measurements per cluster is 24. In comparison, for NGC~5139, NGC~104, and NGC~5904, the number of cluster members with measured RVs in \gaia\ are 1696, 1501, and 194, respectively. Conversely, all clusters in S19 have a minimum of 50 stars with RVs, some containing up to 2500. This contrasts strongly with what can be obtained using integral field spectrographs (IFS). For example, using MUSE, \citet{2018MNRAS.473.5591K} was able to measure up to 20,000 RVs per GC, although at much smaller angular scales than those accessible with \gaia.  

An instrument such as the WEAVE\footnote{William Herschel Telescope Enhanced Area Velocity Explorer} telescope \citep{2024MNRAS.530.2688J} could incorporate both functionalities. WEAVE has a multi-object spectrograph (MOS) with a field of view of 2 square degrees and an IFS of 90x78 arcsec$^2$  (almost twice larger than the MUSE IFS). Together, these components would enable IFS-based studies of the innermost cores of northern GCs complemented by the MOS at larger separations. The proposed wide-field spectroscopic telescope \citep{2024arXiv240305398M} would simultaneously probe much larger core areas ($3x3$\,arcmin$^2$) using its IFS and more than 20\,000 stars in the outer regions of GCs in the southern hemisphere.

The third \gaia\ data release  is the result of the analysis of 34 months of data, and it was limited to measuring RVs for stars with $G_{RVs}<14$ mag. The next data release will be based on 66 months, and this will allow the \gaia\ collaboration to push this limit to $G_{RVs} = 16$~mag \citep{2023A&A...674A...5K}.  By using the photometric transformations from $G$ to $G_{RVs}$ provided by \citet{2023A&A...674A...6S} and examining the cluster members with membership probabilities above $90\%$ in the sample used in this paper, we can establish that $\sim 70,000$ new stars will fall within the \gaia\ observable magnitude range due to the increased depth of observed magnitudes. In \gaia\ DR3, $\sim 63\%$ of all stars in GCs within the \gaia\ observable magnitude range have measured RVs. If this trend persists in \gaia\ DR4, it would imply the presence of $\sim 45,000$ new stars with measured RVs.

The improvements in the astrometry and the increase in the number of stars with measured RVs in \gaia\ DR4 may facilitate the detection of rotation in a larger number of GCs. Based on the results of this study, we conclude that $\sim 200$ individual RVs uniformly distributed throughout a cluster are necessary to measure its rotation, assuming their rotation amplitude is $\sim 6 \, \mathrm{km\,s^{-1}}$. 

\section{Summary}\label{sec:summary}
Our study showed the potential of \gaia\ of measuring the geometry (ellipticity) and rotation of GCs. \gaia\ astrometry allows us to measure the geometry of GCs in regions of the sky where discerning cluster members from background or foreground stars was once challenging, for example, for clusters located in the Galactic plane. It is worth mentioning that astrometry is also useful for studying the outskirts of clusters in greater detail because in these areas, the field contamination has a more significant impact on the measurement of ellipticities than at the core of GCs. The clusters BH~140 and FSR~1758 are notable examples of the \gaia\ capabilities, as we have measured their ellipticities for the first time. This was previously impossible due to their location within high-density star regions. 

Our results are consistent with those previously found using the \hst\, but they were determined for a sample of GCs that was 15 times larger. Our dataset represents the most extensive collection of ellipticities for GCs in the literature. However, it is essential to note that for clusters located at large distances from us, the detection of member stars becomes increasingly challenging because the uncertainties associated with the astrometry increase. Consequently, the number of stars used in the analysis decreases with distance, which in turn increases the statistical uncertainties on the ellipticities. 

By comparing the orientation of the minor axes here with measurements of the projected on-sky rotation axis from the literature, we demonstrated that depending on the reference used for comparison, they are aligned for $76\%$ to $100\%$ of the GCs. The lack of an alignment might arise because the ellipticities and rotation were measured at different angular scales. Using the RVs from \gaia, we have confirmed the rotation of NGC~104, NGC~5139, and NGC~5904. For these clusters, we found that their geometry is aligned with the orientation of the rotation.

The fourth data release of \gaia\ will provide RVs for stars up to two magnitudes fainter than in the third realese. Together with an additional 32 months of observations, this will enhance the ability of \gaia\ to measure the rotation of GCs.

\begin{acknowledgements}
This research received support from the European Research Council (ERC) under the European Union's Horizon 2020 research and innovation programme (Grant Agreement No. 947660). RIA is funded by the SNSF through a Swiss National Science Foundation Eccellenza Professorial Fellowship (award PCEFP2\_194638). 

We thank Henry Leung for designing the \href{https://milkyway-plot.readthedocs.io/en/stable/}{milkyway-plot} python package used to generate Fig. \ref{fig:mw_ellipticty}.

This work has made use of data from the European Space Agency (ESA) mission
{\it Gaia} (\url{https://www.cosmos.esa.int/gaia}), processed by the {\it Gaia}
Data Processing and Analysis Consortium (DPAC,
\url{https://www.cosmos.esa.int/web/gaia/dpac/consortium}). Funding for the DPAC
has been provided by national institutions, in particular the institutions
participating in the {\it Gaia} Multilateral Agreement. 

This research has made use of NASA's Astrophysics Data System; the SIMBAD database and the VizieR catalog access tool\footnote{\url{http://cdsweb.u-strasbg.fr/}} provided by CDS, Strasbourg; Astropy\footnote{\url{http://www.astropy.org}}, a community-developed core Python package for Astronomy \citep{astropy:2013, astropy:2018}; TOPCAT\footnote{\url{http://www.star.bristol.ac.uk/~mbt/topcat/}} \citep{2005ASPC..347...29T}.

\end{acknowledgements}

\bibliographystyle{aa}
\bibliography{refs}

\begin{appendix}

\section{Cluster parameters}\label{sec:parameters}
In this appendix, we compile the cluster parameters discussed in this paper, covering ellipticities in Table \ref{tab:ellipticities}, the median axial ratio for the cluster in our sample and literature in Table \ref{tab:median_ratios} and position angles in Table \ref{tab:position_angles}.

\begin{table}[!htp]
  \centering
  \caption{Ellipticity of GCs}
    \setlength{\tabcolsep}{0.08cm}
\begin{tabular}{lcccccccccc}\toprule  
Cluster  & $\epsilon$ & $\epsilon_{\mathrm{C10}}$ & $\epsilon_{\mathrm{H10}}$ & $\epsilon_{\mathrm{F14}}$ & $\Theta$ & N$_{\mathrm{PCA}}$ & Photometric $\Theta_{F14}$ & N$_{\mathrm{F14}}$  &  $\Theta_{\mathrm{M23}}$ & N$_{\mathrm{M23}}$  \\   \midrule
NGC 5139 & $0.064 \pm 0.002$ & $0.210 \pm 0.020$ & 0.17 &  & 0.671 & 147516 &  &  &  &  \\
NGC 104  & $0.059 \pm 0.004$ & $0.160 \pm 0.020$ & 0.09 &  & 0.668 & 102838 &  &  & 0.064 & 28588 \\
NGC 6752 & $0.023 \pm 0.006$ & $0.040 \pm 0.020$ & 0.04 &  & 0.417 & 38979 &  &  & 0.040 & 13721 \\   
NGC 6205 & $0.050 \pm 0.007$ & $0.120 \pm 0.020$ & 0.11 & $0.018 \pm 0.011$ & 0.416 & 28204 & 0.043 & 152392 &  &  \\
... & ... & ... & ... & ... & ... & ... & ... & ... & ... & ... \\

              \bottomrule
\end{tabular}
\tablefoot{The complete version of this table is available at the CDS. To estimate the ellipticity, we consistently applied the definition  $\epsilon = 1 - \sqrt{\lambda_{b}/\lambda_{a}} = 1 - b/a $. The columns $\Theta$, $\Theta_{F14}$, and $\Theta_{M23}$ represent the angular scale used to measure the ellipticities in their respective papers, measured in degrees. This scale is determined by measuring the angular separation in degrees between the cluster's center and the most distant star in the sample. This column was not determined for the H10 and C10 samples as the data are not available. For the same reason, the $\Theta_{14}$ column for NGC~6626 is left empty. The columns  N$_{\mathrm{PCA}}$, N$_{\mathrm{F14}}$, and  N$_{\mathrm{M23}}$ represent the number of stars in each sample.    }
    \label{tab:ellipticities}
\end{table}

\begingroup 
\renewcommand{\arraystretch}{1.5}
\begin{table}[!htp]
    \centering
    \caption{Comparison of the median axial ratio from this paper with the literature values.}
    \begin{tabular}{@{}ccccc}\toprule
          $b/a$ from this paper  &  $b/a$  from the literature & Reference & Number of clusters  \\ \midrule
          
          $0.951^{+0.021}_{-0.036}$ & $0.940^{+0.050}_{-0.090}$ &  H10 &  100 \\
          
          $0.948^{+0.022}_{-0.046}$ & $0.865^{+0.065}_{-0.095}$ & C10 &  116 \\
          
          $0.964^{+0.011}_{-0.017}$ & $0.986^{+0.009}_{-0.004}$ & F14 &  11 \\
          
          $0.935^{+0.033}_{-0.090}$ &  &  &  163  \\
         \bottomrule
    \end{tabular}
    \tablefoot{Median value, 16th and 84th percentiles of the distribution of $b/a$.  }
    \label{tab:median_ratios}
\end{table}
\endgroup

\begin{table}[!htp]
  \centering
  \caption{Position angles of ellipses fitted to GC member stars}
  \setlength{\tabcolsep}{0.1cm}
  \renewcommand{\arraystretch}{1.2}
\begin{tabular}{lcccccc}\toprule  
Cluster  &         $\theta$  & Kinematic $\theta_{\mathrm{F14}}^{T}$   &  Photometric $\theta_{\mathrm{F14}}^{T}$  &   $\theta_{\mathrm{S19}}^{T}$ &   $\theta_{\mathrm{M23}}^{T}$ \\   \midrule
NGC 5139 & $ 164.0 \pm 1.1$  &                     &                   &  $ 170.2 \pm 7.6$&                     \\
NGC 104 & $ 46.7 \pm 1.7$    &                      &                   &  $44.3 \pm 4.6$ &  $45.9^{1.7}_{1.7}$      \\
NGC 6752& $ 155.7 \pm 12.4$  &                     &                   &                 & $71.2^{+17.2}_{-17.2}$   \\   
NGC 6205 & $ 139.4	\pm 4.5$ &      $163.5 \pm 8.6$ & $ 154.2 \pm 17.2$ &   $ 165.5\pm 14.2$ &    \\
...      & ...               &   ...     &  ...  &      ... &            ...           \\
              \bottomrule
\end{tabular}
\tablefoot{The complete version of this table is available at the CDS. All angles are measured in degrees, they have been transformed to the same reference system and are defined from $0^{\circ}$ to $180^{\circ}$.   }

    \label{tab:position_angles}
\end{table}

\end{appendix}

\end{document}